\documentclass{iopart}
\usepackage{graphicx}
\usepackage{hyperref}
\usepackage{bbm,amssymb,amsbsy}

\usepackage{color}

\newcommand{\R}{\mathbbm{R}}

\newcommand{\rr}{\mathbbm{R}}

\newcommand{\id}{\mathbbm{1}}

\newcommand{\diag}{\,{\rm diag}\,}

\newcommand{\sy}[1]{Sp_{(#1,\R)}}

\renewcommand{\tr}{{\rm Tr}\,}
\renewcommand{\det}{{\rm Det}\,}

\newcommand{\be}{\begin{equation}}
\newcommand{\ee}{\end{equation}}
\newcommand{\bea}{\begin{eqnarray}}
\newcommand{\eea}{\end{eqnarray}}
\newcommand{\ket}[1]{|#1\rangle}
\newcommand{\bra}[1]{\langle#1|}

\newcommand{\N}{{\cal N}}

\newcommand{\eq}[1]{Eq.~(\ref{#1})}

\newcommand{\ie}{\emph{i.e.}~}

\begin{document}

\title[Entanglement of two-mode Gaussian states: characterization and experimental \ldots]
{Entanglement of two-mode Gaussian states: characterization and
experimental \\ production and manipulation}

\author{Julien Laurat$\dag$, Ga{\"e}lle Keller$\dag$, Jos\'e Augusto
Oliveira-Huguenin$\dag$, Claude Fabre$\dag$, and \\ Thomas
Coudreau$\dag\ddag$\footnote[3]{To whom correspondence should be
addressed (coudreau@spectro.jussieu.fr)}}

\address{$\dag$ Laboratoire Kastler Brossel, UPMC, Case 74, 4 Place Jussieu, 75252 Paris cedex
05, France}

\address{$\ddag$ Laboratoire Mat{\'e}riaux et Ph{\'e}nom{\`e}nes Quantiques,
Case 7021, Universit{\'e} Denis Diderot, 2 Place Jussieu, 75251 Paris
cedex 05, France}

\author{Alessio
Serafini$\natural\sharp$, Gerardo Adesso$\natural$, and \\ Fabrizio
Illuminati$\natural$}

\address{$\natural$ Dipartimento di Fisica ``E. R. Caianiello'',
Universit\`a di Salerno; CNR-Coherentia, Gruppo di Salerno; and
INFN Sezione di Napoli-Gruppo Collegato di Salerno, Via S. Allende,
84081 Baronissi (SA), Italy}

\address{$\sharp$ Department of Physics \& Astronomy, University College London,
Gower Street, London WC1E 6BT, United Kingdom}

\begin{abstract}
A powerful theoretical structure has emerged in recent years on
the characterization and quantification of entanglement in
continuous-variable systems. After reviewing this framework, we
will illustrate it with an original set-up based on a type-II OPO
with adjustable mode coupling. Experimental results allow a direct
verification of many theoretical predictions and provide a sharp
insight into the general properties of two-mode Gaussian states
and entanglement resource manipulation.
\end{abstract}

\pacs{03.67.Mn, 42.65.Yj, 42.50.Dv, 42.50.Lc}

\date{June 13, 2005}

\maketitle

\section{Introduction}

Quantum information aims at the treatment and transport of
information using the laws of quantum physics. For these goals,
continuous variables (CV) of the electromagnetic field have
emerged as a powerful tool \cite{CV,brareview,cvqipbook}. In this
context, entanglement is an essential ressource. The purpose of
this paper is to link the very powerful mathematical description
of gaussian states based on covariance matrices and the
experimental production and manipulation of entanglement.

Experimentally, entanglement can be obtained directly by type-II
parametric interaction deamplifying either the vacuum fluctuations
as was demonstrated in the seminal experiment by Ou \emph{et al.}
\cite{ou92} (or in recent experiments \cite{bf,saopaulo}) or the
fluctuations of a weak injected beam \cite{eprtaiyuan}. It can
also be obtained indirectly by mixing on a beam splitter two
independent squeezed beams. The required squeezing can be produced
by Kerr effects -- using optical fibers \cite{erlangen} or cold
atoms in an optical cavity \cite{josse} -- or by type-I parametric
interaction in a cavity \cite{wu86,australie}. Single-pass type-I
interaction in a non-colinear configuration can also generate
directly entangled beams as demonstrated recently by Wenger
\emph{et al.} in the pulsed regime \cite{wengerEPR}. All these
methods produce a symmetric entangled state enabling dense coding,
the teleportation of coherent
\cite{teleportationkimble,australie,furusawa} or squeezed states
\cite{teleportationfurusawa} or entanglement swapping
\cite{swappingtaiyuan,furusawa}. These experiments generate an
entangled state with a covariance matrix in the so-called
`standard form' \cite{duan00,simon00}, without having to apply any
local linear unitary transformations such as beam-splitting or
phase-shifts to exploit it optimally in quantum information
protocols.

However, it has been recently shown in \cite{eprparis} that, when
a birefringent plate is inserted inside the cavity of a type-II
optical parametric oscillator, \emph{i.e.} when mode coupling is
added, the generated two-mode state remains symmetric but
entanglement is not observed on orthogonal quadratures: the state
produced is not in the standard form. The entanglement of the two
emitted modes in this configuration is not optimal: it is indeed
possible by passive non-local operations to select modes that are
more entangled. Our original system provides thus a good insight
into the quantification and manipulation of the entanglement
resources of two-mode Gaussian states. In particular, as just
anticipated, it allows to confirm experimentally the theoretical
predictions on the entangling capacity of passive optical elements
and on the selection of the optimally entangled bosonic modes
\cite{passive}.

The paper is organized as follows: we start by giving a general
overview of gaussian states and defining the covariance matrix
formalism, focusing in particular on two-mode states. Over the
last years, a great deal of attention has been devoted to defining
not only criterion of entanglement but also quantification of this
entanglement. Section \ref{ment} focus on different such measures
which are interpreted in the covariance matrix formalism. Effects
of mode coupling on the generated entanglement is then discussed
in detail (section \ref{sec:modecoupling}). The experimental setup
used to generate and quantify the entanglement is presented in
section \ref{sec:setup}. In section \ref{sec:expmesent},
experimental measures of entanglement are realized and we discuss
on the effect of noise of the covariance matrix on the
entanglement measures. Finally, we show how optimization of the
resource is obtained by a passive operation -- namely a
polarization adjustment -- operated on the two-mode state (section
\ref{sec:nonsf}).


\section{Gaussian states: general overview\label{sec2m}}
A continuous variable (CV) system is described by a Hilbert space
${\cal H}=\bigotimes_{i=1}^{n} {\cal H}_{i}$ resulting from the
tensor product of infinite dimensional Fock spaces ${\cal
H}_{i}$'s. Let $a_{i}$ and $a_i^\dag$ be the annihilation and
creation operators acting on ${\cal H}_{i}$ (ladder operators),
and $\hat x_{i}=(a_{i}+a^{\dag}_{i})$ and $\hat
p_{i}=(a_{i}-a^{\dag}_{i})/i$ be the related quadrature phase
operators. The corresponding phase space variables will be denoted
by $x_{i}$ and $p_{i}$. Let $\hat X = (\hat x_{1},\hat
p_{1},\ldots,\hat x_{n},\hat p_{n})$ denote the vector of the
operators $\hat x_{i}$ and $\hat p_{i}$. The canonical commutation
relations for the $\hat X_{i}$ can be expressed in terms of the
symplectic form ${\Omega}$
\[
[\hat X_{i},\hat X_j]=2i\Omega_{ij} \; ,
\]
\[
{\rm with}\quad{\Omega}\equiv \bigoplus_{i=1}^{n}
{\omega}\; , \quad {\omega}\equiv \left( \begin{array}{cc}
0&1\\
-1&0
\end{array}\right) \; .
\]

The state of a CV system can be equivalently described by a
positive trace-class operator (the density matrix $\varrho$) or by
quasi--probability distributions. In the following, we shall adopt
the Wigner quasi--probability function $W(R)$ defined, for any
density matrix, as the Fourier transform of the symmetrically
ordered characteristic function \cite{barnett}:
$$
W(R) \equiv \frac{1}{\pi^2}\int_{\rr^{2n}} \tr{[\varrho D_Y]} \,{\rm e}^{i Y^T \Omega R}
\,{\rm d}^{2n}Y \, ,\,\,\,  R \in \rr^{2n} \, ,
$$
where the displacement operators $D_Y$ (describing the effect on the
field of a classical driving current) are defined as
\begin{equation}
    D_Y = e^{ i Y^T \Omega \hat X}, \,\,\, Y \in \rr^{2n} \, .
\end{equation}
The $2n$-dimensional space of definition of the Wigner function,
associated to the quadratic form $\Omega$, is referred to as phase
space, in analogy with classical Hamiltonian dynamics. In Wigner
phase space picture, the tensor product ${\cal H}=\bigotimes{\cal
H}_{i}$ of the Hilbert spaces ${\cal H}_{i}$'s of the $n$ modes
results in the direct sum $\Gamma=\bigoplus\Gamma_{i}$ of the
phase spaces $\Gamma_{i}$'s.

States with Gaussian characteristic functions and
quasi--probability distributions are referred to as Gaussian
states. Such states are at the heart of information processing in
CV systems \cite{brareview} and are the subject of our analysis.
By definition, a Gaussian state $\varrho$ is completely
characterised by the first and second statistical moments of the
field operators, which will be denoted, respectively, by the
vector of first moments $\bar X\equiv\left(\langle\hat X_{1}
\rangle,\langle\hat X_{2}\rangle,\ldots,\langle\hat
X_{2n-1}\rangle, \langle\hat X_{2n}\rangle\right)$ and the
covariance matrix (CM) $\Gamma$ of elements
\begin{equation}
\Gamma_{ij}\equiv\frac{1}{2}\langle \hat{X}_i \hat{X}_j +
\hat{X}_j \hat{X}_i \rangle -
\langle \hat{X}_i \rangle \langle \hat{X}_j \rangle \, , \label{covariance}
\end{equation}
where, for any observable $\hat{o}$, the expectation value
$\langle\hat o\rangle\equiv\,{\rm Tr}(\varrho\hat o)$. Notice
that, according to our definition of the quadrature operators in
terms of the ladder operators, the entries of the CM are real
numbers. Being the variances and covariances of quantum operators,
such entries are obtained by noise variance and noise correlation
measurements. They can be expressed as energies by multiplying
them by the quantity $\hbar \omega$, where $\omega$ is the
frequency of the considered mode. In fact, for any $n$-mode state
the quantity $\hbar \omega\tr({\Gamma}/4)$ is just the
contribution of the second moments to the average of the ``free''
Hamiltonian $\sum_{i=1}^{n} (a^{\dag}_ia_i + 1/2)$.

Coherent states, resulting from the application of displacement
operators $D_X$ to the vacuum state $\ket{0}$, are Gaussian states
with CM $\Gamma=\id$ and first statistical moments $\bar X = X$.
First moments can be arbitrarily adjusted by local unitary
operations (displacements), which cannot affect any property
related to entropy or entanglement. In the present experimental
case, the  fields are produced by an OPO below the oscillation
threshold, and the means are actually zero. Let us note that this
can be done without any loss of generality.

The canonical commutation relations and the positivity of the
density matrix $\varrho$ imply
\begin{equation}
\Gamma+ i\Omega\ge 0 \; ,
\label{bonfide}
\end{equation}
meaning that all the eigenvalues of the (hermitian) matrix
$\Gamma+i\Omega$ have to be greater or equal than zero. Inequality
(\ref{bonfide}) is the necessary and sufficient constraint the
matrix $\Gamma$ has to fulfill to be a CM corresponding to a
physical Gaussian state \cite{simon87,simon}. Note that the previous
condition is necessary for the CM of {\em any} (generally non
Gaussian) state, as it generalises to many modes the
Robertson-Schr\"odinger uncertainty relation, reducing to the
familiar Heisenberg principle for pure, uncorrelated states. We
mention that such a constraint implies $\Gamma\ge0$.


A major role in the theoretical and experimental manipulation of
Gaussian states is played by unitary operations which preserve the
Gaussian character of the states on which they act. Such operations
are all those generated by terms of the first and second order in
the field operators. So, beside the already discussed displacements,
also the unitary operations generated by terms of the second order
are Gaussian. As a consequence of the Stone-Von Neumann theorem, any
such operation at the Hilbert space level corresponds, in phase
space, to a symplectic transformation, \emph{i.e.} to a linear
transformation $S$ which preserves the symplectic form $\Omega$, so
that $\Omega=S^T \Omega S$, \emph{i.e.} it preserves the commutators
between the different operators. Symplectic transformations on a
$2n$-dimensional phase space form the (real) symplectic group,
denoted by $\sy{2n}$. Such transformations act linearly on first
moments and ``by congruence'' on covariance matrices ({\ie}so that
$\Gamma\mapsto S^{T} \Gamma S$). One has $\det{S}=1$,
$\forall\,S\in\sy{2n}$.

Ideal beam splitters, phase shifters and squeezers are described
by symplectic transformations. In fact single and two--mode
squeezings occurring, respectively, in degenerate and non
degenerate parametric down conversions, are described by the
operators \be U_{ij,r,\varphi}=\,{\rm e}^{\frac12 (\varepsilon
a_i^{\dag}a_j^{\dag} -\varepsilon^{*}a_ia_j)}  \quad {\rm
with}\;\; \varepsilon=r\,{\rm e}^{i2\varphi}\, , r \in \R, \varphi
\in [0,2\pi], \label{opersq}\ee resulting in single-mode squeezing
of mode $i$ for $i=j$. The representation in phase space of the
operation $U_{ij,r,\varphi}$ for $i\neq j$ is given by the linear
transformation $S_{ij,r,\varphi}$ \be
S_{ij,r,\varphi}=\left(\begin{array}{cccc}
c-hs&0&ks&0\\
0&c+hs&0&-ks\\
ks&0&c+hs&0\\
0&-ks&0&c-hs
\end{array}\right) \quad {\rm for}\,i\neq j \, ,
\ee where $c=\cosh(2r)$, $s=\sinh(2r)$, $h=\cos(2\varphi)$,
$k=\sin(2\varphi)$ ($r$ and $\varphi$ are same as in \eq{opersq})
and the matrix is understood to act on the couple of modes $i$ and
$j$. Beam splitters are described by the operators \be
O_{ij,\theta}= \,{\rm e}^{\theta a_i^{\dag}a_j-\theta
a_ia_j^{\dag}} \; ,\theta \in [0,2\pi], \label{operbs}\ee
corresponding to symplectic rotations $R_{ij,\theta}$ in phase
space \be R_{ij,\theta}=\left(\begin{array}{cccc}
\cos(\theta)&0&-\sin(\theta)&0\\
0&\cos(\theta)&0&-\sin(\theta)\\
\sin(\theta)&0&\cos(\theta)&0\\
0&\sin(\theta)&0&\cos(\theta)
\end{array}\right) \quad {\rm for}\, i\neq j \, .\label{rota}
\ee The angle $\theta$ is defined by \eq{operbs} and, again, the
matrix is understood to act on the $i$ and $j$ modes. Single-mode
symplectic operations are easily retrieved as well, being just
combinations of single mode (two dimensional) rotations and of
single mode squeezings of the form $\diag(\,{\rm e}^{r},\,{\rm
e}^{-r})$ for $r>0$. Now, symplectic transformations in phase
space are generated by exponentiation of matrices written as
$J\Omega$, where $J$ is antisymmetric \cite{pramana}. Such
generators can be symmetric or antisymmetric. The operations
$R_{ij,\theta}$, generated by antisymmetric operators are
orthogonal and, acting by congruence on the CM $\Gamma$, preserve
the value of $\tr{\Gamma}$. Since $\tr{\Gamma}$ gives the
contribution of the second moments to the average of the
Hamiltonian $\bigoplus_i a_i^{\dag}a_i$, these transformations are
said to be `passive', or `energy preserving' (they belong to the
compact subgroup of $\sy{2n}$). Instead, operations
$S_{ij,r,\varphi}$, generated by symmetric operators, are not
orthogonal and do not preserve $\tr{\Gamma}$ (they belong to the
non compact subgroup of $\sy{2n}$). This mathematical difference
between squeezers and phase space rotations accounts, in a quite
elegant way, for the difference between `active' (\ie energy
consuming) and `passive' (\ie energy preserving) optical
transformations.

Finally, let us recall that, due to Williamson theorem
\cite{williamson36}, the CM of a $n$--mode Gaussian state can
always be written as
\begin{equation}
\Gamma=S^T \nu S \; , \label{willia}
\end{equation}
where $S\in Sp_{(2n,\mathbb{R})}$ and $\nu$ is the CM
\begin{equation}
\nu=\,{\rm diag}({\nu}_{1},{\nu}_{1},\ldots,{\nu}_{n},{\nu}_{n}) \, ,
\label{therma}
\end{equation}
corresponding to a tensor product of thermal states with diagonal
density matrix $\varrho^{_\otimes}$ given by \be
\varrho^{_\otimes}=\bigotimes_{i}
\frac{2}{\nu_{i}+1}\sum_{k=0}^{\infty}\left(
\frac{\nu_{i}-1}{\nu_{i}+1}\right)\ket{k}_{i}{}_{i}\bra{k}\; ,
\label{thermas} \ee $\ket{k}_i$ being the $k$-th number state of the
Fock space ${\cal H}_{i}$. The dual (Hilbert space) formulation
of \eq{willia} then reads: $\varrho=U^{\dag}\,\varrho^{_\otimes}\,
U$, for
some unitary $U$.\\
The quantities $\nu_{i}$'s form the symplectic spectrum of the CM
$\Gamma$ and can be computed as the eigenvalues of the matrix
$|i\Omega\Gamma|$ \cite{abs}. Such eigenvalues are in fact
invariant under the action
of symplectic transformations on the matrix $\Gamma$.\\
The symplectic eigenvalues $\nu_{i}$ encode essential informations
on the Gaussian state $\Gamma$ and provide powerful, simple ways
to express its fundamental properties. For instance, in terms of
the symplectic eigenvalues $\nu_{i}$, the uncertainty relation
(\ref{bonfide}) simply reads \be {\nu}_{i}\ge1 \; .
\label{sympheis} \ee

Also the entropic quantities of Gaussian states can be expressed in
terms of their symplectic eigenvalues and invariants. Notably, the
purity $\tr{\varrho^2}$ of a Gaussian state $\varrho$ is simply
given by the symplectic invariant
$\det{\Gamma}=\prod_{i=1}^{n}\nu_i$, being $\tr{\varrho^2} =
(\det{\Gamma})^{-1/2}$ \cite{paris}.

\subsection{Two--mode states}
Since this work is focused on two--mode Gaussian states, we
briefly review here some of their basic properties. The expression
of the two--mode CM $\Gamma$ in terms of the three $2\times 2$
matrices $\alpha$, $\beta$, $\gamma$ will be useful
\begin{equation}
\Gamma\equiv\left(\begin{array}{c|c} {\alpha}&{\gamma}\\
\hline {\gamma}^{T}&{\beta}
\end{array}\right)\, . \label{espre}
\end{equation}
For any two--mode
CM ${\Gamma}$ there exist local
symplectic operations $S_{1}$ and $S_{2}$ (each $S_j$ acting on one of the two modes),
such that their direct sum $S_{l}=S_{1}\oplus S_{2}$ (corresponding to the
tensor product of local unitary operations)
brings
${\Gamma}$ to the so called standard form ${\Gamma}_{sf}$
\cite{simon00, duan00}
\begin{equation}
S_{l}^{T}{\Gamma}S_{l}={\Gamma}_{sf} \equiv
\left(\begin{array}{cc|cc}
a&0&c_{+}&0\\
0&a&0&c_{-}\\ \hline
c_{+}&0&b&0\\
0&c_{-}&0&b
\end{array}\right)\; . \label{stform}
\end{equation}
States whose standard form fulfills $a=b$ are said to be symmetric.
Let us recall that any pure state is symmetric and fulfills
$c_{+}=-c_{-}=\sqrt{a^2-1}$. The correlations $a$, $b$, $c_{+}$, and
$c_{-}$ are determined by the four local symplectic invariants ${\rm
Det}{\Gamma}=(ab-c_{+}^2)(ab-c_{-}^2)$, ${\rm Det}{\alpha}=a^2$,
${\rm Det}{\beta}=b^2$, ${\rm Det}{\gamma}=c_{+}c_{-}$. Therefore,
the standard form corresponding to any CM is unique (up to a common
sign flip in $c_-$ and $c_{+}$).

For two--mode states, the uncertainty principle
Ineq.~(\ref{bonfide}) can be recast as a constraint on the
$Sp_{(4,{\mathbb R})}$ invariants ${\rm Det}\Gamma$ and
$\Delta(\Gamma)={\rm Det}{\alpha}+\,{\rm Det}{\beta}+2 \,{\rm
Det}{\gamma}$:
\begin{equation}
\Delta(\Gamma)\le1+\,{\rm Det}\Gamma
\label{sepcomp}\; .
\end{equation}

The symplectic eigenvalues of a two--mode Gaussian state will be
named $\nu_{-}$ and $\nu_{+}$, with $\nu_{-}\le \nu_{+}$, with the
Heisenberg uncertainty relation reducing to \be \nu_{-}\ge 1 \; .
\ee A simple expression for the $\nu_{\mp}$ can be found in terms
of the two $Sp_{(4,\mathbb{R})}$ invariants (invariants under
global, two--mode symplectic operations) \cite{logneg, serafozzi}
\begin{equation}
2{\nu}_{\mp}^2=\Delta(\Gamma)\mp\sqrt{\Delta(\Gamma)^2
-4\,{\rm Det}\,\Gamma} \, . \label{sympeig}
\end{equation}

A subclass of Gaussian states with a major interest in
experimental quantum optics and in the practical realization of CV
quantum information protocols is constituted by the nonsymmetric
two--mode squeezed thermal states. Let $S_{r}=S_{12,r,\pi/4}$ be
the two mode squeezing operator with real squeezing parameter $r$,
and let $\varrho^{_\otimes}_{\nu_i}$ be a tensor product of
thermal states with CM ${\nu}_{\nu_{\mp}}= {\mathbbm 1}_{1}\nu_-
\oplus {\mathbbm 1}_{2}\nu_{+}$, where $\nu_{\mp}$ is the
symplectic spectrum of the state. A nonsymmetric two-mode squeezed
thermal state $\xi_{\nu_{i},r}$ is defined as
${\xi}_{\nu_{i},r}=S_{r}\varrho^{_\otimes}_{\nu_i}S_{r}^{\dag}$,
corresponding to a standard form with \bea
a&=&{\nu_-}\cosh^{2}r+{\nu_{+}}\sinh^{2}r \; ,\nonumber\\
b&=&{\nu_{-}}\sinh^{2}r+{\nu_{+}}\cosh^{2}r \; ,\\
c_{\pm}&=&\pm\frac{\nu_-+\nu_+}{2}\sinh2r \; . \nonumber
\eea
In the symmetric instance (with $\nu_-=\nu_+=\nu$)
these states reduce to two--mode squeezed thermal
states. The covariance matrices of these states are
symmetric standard forms with
\be
a=\nu\,{\cosh2r}\; , \quad
c_{\pm}=\pm\nu\,{\sinh2r} \, . \label{sqthe}
\ee
These are the states which occur in most realistic parametric down conversion
processes, like the one which will be discussed later on in the paper.
In the pure case, for which $\nu=1$, one recovers the
two--mode squeezed vacua. Such states encompass
all the standard forms associated to pure states: any two--mode
Gaussian state can reduced to a squeezed vacuum by means of
unitary local operations.

\section{Entanglement of Gaussian states\label{ment}}

This section aims at reviewing the main results on the
qualification and quantification of entanglement for Gaussian
states of CV systems, which will be exploited in the following.

The positivity of the partially transposed state (PPT criterion) is
necessary and sufficient for the separability of two--mode Gaussian
states \cite{simon00} (and, more generally, of all $(1+n)$--mode
Gaussian states under $1\times n$-mode bipartitions \cite{werwolf}
and of bisymmetric $(m+n)$--mode Gaussian states under $m\times
n$-mode bipartitions \cite{unitarily}). In general, the partial
transposition $\tilde{\varrho}$ of a bipartite quantum state
$\varrho$ is defined as the result of the transposition performed on
only one of the two subsystems in some given basis. It can be
promptly inferred from the definition of the Wigner function $W(X)$
that the action of partial transposition amounts, in phase space, to
a mirror reflection of one of the four canonical variables. In terms
of $Sp_{(2,\mathbb{R})}\oplus Sp_{(2,\mathbb{R})}$ invariants, this
reduces to a sign flip in ${\rm Det}\,\gamma$. Therefore the
invariant $\Delta(\Gamma)$ is changed into $\tilde{\Delta}({\Gamma})
=\Delta(\tilde{\Gamma})=\,{\rm Det}\,{\alpha}+ \,{\rm
Det}\,{\beta}-2\,{\rm Det}\,{\gamma}$. Now, the symplectic
eigenvalues $\tilde{\nu}_{\mp}$ of $\tilde{\Gamma}$ read \be
\tilde{\nu}_{\mp}=
\sqrt{\frac{\tilde{\Delta}(\Gamma)\mp\sqrt{\tilde{\Delta}(\Gamma)^2
-4\,{\rm Det}\,\Gamma}}{2}} \, . \label{sympareig} \ee The PPT
criterion thus reduces to a simple inequality that must be satisfied
by the smallest symplectic eigenvalue $\tilde{\nu}_{-}$ of the
partially transposed state \be \tilde{\nu}_{-}\ge 1 \: ,
\label{symppt} \ee which is equivalent to \be
\tilde{\Delta}(\Gamma)\le \,{\rm Det}\,\Gamma+1 \; . \label{ppt} \ee
The above inequalities imply ${\rm Det}\,{\gamma}=c_{+}c_{-}<0$ as a
necessary condition for a two--mode Gaussian state to be entangled.
Therefore, the quantity $\tilde{\nu}_{-}$ encodes all the
qualitative characterization of the entanglement for arbitrary (pure
or mixed) two--modes Gaussian states.

Let us now briefly focus on the entanglement qualification of
symmetric states, which will be the subject of the experimental
investigations presented in the paper. It is immediately apparent
that, because $a=b$, the partially transposed CM in standard form
$\tilde{\Gamma}$ (obtained by flipping the sign of $c_-$) is
diagonalized by the orthogonal, symplectic transformation
$R_{12,\pi/4}$ of \eq{rota}, resulting in a diagonal matrix with
entries $a\mp |c_\mp|$. The symplectic eigenvalues of such a matrix
are then easily retrieved by applying local squeezings. In
particular, the smallest eigenvalue $\tilde{\nu}_{-}$ is simply
given by \be \tilde{\nu}_{-}=\sqrt{(a-|c_{+}|)(a-|c_{-}|)} \; .
\label{symeig} \ee Note that also the original standard CM $\Gamma$
with $a=b$ could be diagonalized ({\em not symplectically}, since
the four diagonal entries are generally all different) by the same
beam splitter transformation $R_{12,\pi/4}$, with the same
orthogonal eigenvalues $a\mp |c_\mp|$. It is immediate to verify
that $\tilde{\nu}_-$ is just given by the geometric average between
the two smallest of such orthogonal eigenvalues of $\Gamma$. The two
quadratures resulting from the previous beam splitter transformation
select orthogonal directions in phase space with respect to the
original ones, so they will be referred to as `orthogonal'
quadratures. Notice that, in the experimental practice, this allows
for the determination of the entanglement through the measurement of
diagonal entries (noise variances) of the CM after the application
of a balanced beam splitter (which embodies the transformation
$R_{12,\pi/4}$).

To explore further consequences of this fact, let us briefly
recall some theoretical results on the generation of entanglement
under passive (energy-preserving) transformations, which will be
precious in the following. As shown in Ref.~\cite{passive}, the
minimum value for $\tilde{\nu}_{-}$ ({\ie}the maximal
entanglement) attainable by passive transformations is given by
$\tilde{\nu}_{-}^{2}=\lambda_1\lambda_2$, where $\lambda_1$ and
$\lambda_2$ are the two smallest eigenvalues of $\Gamma$.
Therefore, the entanglement of symmetric states {\em in standard
form} cannot be increased through energy preserving operations,
like beam splitter and phase shifters. On the other hand, as it
will be carefully discussed in great detail in the next section,
the insertion of a birefringent plate in a type-II optical
parametric oscillator results in states symmetric but not in
standard form. In such a case the entanglement can be optimized by
the action of a (passive) phase shifter.

A measure of entanglement which can be computed for general
Gaussian states is provided by the negativity $\N$, first
introduced in Refs.~\cite{zircone,eisert}, later thoroughly
discussed and extended in Ref.~\cite{logneg} to CV systems. The
negativity of a quantum state $\varrho$ is defined as \be {\cal
N}(\varrho)=\frac{\|\tilde \varrho \|_1-1}{2}\: , \ee where
$\tilde\varrho$ is the partially transposed density matrix and
$\|\hat o\|_1=\,{\rm Tr}|\hat o|$ stands for the trace norm of the
hermitian operator $\hat o$. The quantity ${\cal N} (\varrho)$ is
equal to $|\sum_{i}\lambda_{i}|$, the modulus of the sum of the
negative eigenvalues of $\tilde\varrho$, quantifying the extent to
which $\tilde\varrho$ fails to be positive. Strictly related to
$\N$ is the logarithmic negativity $E_{\N}$, defined as
$E_{\N}\equiv \log_2 \|\tilde{\varrho}\|_{1}$, which constitutes
an upper bound to the {\em distillable entanglement} of the
quantum state $\varrho$. Both the negativity and the logarithmic
negativity have been proven to possess the crucial property of
being monotone under LOCC (local operations and classical
communications) \cite{logneg,plenio05}.

For any two--mode Gaussian state $\varrho$ the negativity is a
simple decreasing function of $\tilde{\nu}_{-}$, which is thus
itself an inverse quantifier of entanglement: \be
\|\tilde\varrho\|_{1}=\frac{1}{\tilde{\nu}_{-}}\;\Rightarrow
\N(\varrho)=\max \, \left[
0,\frac{1-\tilde{\nu}_{-}}{2\tilde{\nu}_{-}} \right] \, , \ee \be
E_{\N}(\varrho)=\max\,\left[0,-\log_2 \tilde{\nu}_{-}\right] \, .
\ee This expression quantifies the amount by which PPT inequality
(\ref{symppt}) is violated. The symplectic eigenvalue
$\tilde{\nu}_{-}$ thus completely qualifies and quantifies the
quantum entanglement of a Gaussian state $\Gamma$.

For symmetric Gaussian states one can also compute \cite{giedke03}
the entanglement of formation $E_{F}$ \cite{bennet96}. We recall
that the entanglement of formation $E_{F}$ of a quantum state
$\varrho$ is defined as \be
E_{F}(\varrho)=\min_{\{p_i,\ket{\psi_i}\}}\sum_i p_i
E(\ket{\psi_i}) \; , \label{eof} \ee where the minimum is taken
over all the pure states realizations of $\varrho$:
\[
\varrho=\sum_i p_i \ket{\psi_i}\bra{\psi_i} \; .
\]
The quantity $E_F$ satisfies all the requirements of a proper
entanglement measure. The asymptotic regularization of the
entanglement of formation is equal to the entanglement cost $E_C
(\varrho)$, defined as the minimum number of singlets (maximally
entangled antisymmetric two-qubit states) which is needed to prepare
the state $\varrho$ through local operations and classical
communication \cite{ecost}. In formulae:
\be
E_C(\varrho)=\lim_{n\rightarrow\infty} \frac{E_F(\varrho^{\otimes n})}{n} \; .
\label{ecost}
\ee

The optimal convex decomposition of \eq{eof} can be found for
symmetric states and turns out to be Gaussian, allowing for the
determination of the entanglement of formation $E_F$: \be E_F =
\max\left[ 0,h(\tilde{\nu}_{-}) \right] \; , \label{eofgau} \ee with
\[
h(x)=\frac{(1+x)^2}{4x}\log_2 \left(\frac{(1+x)^2}{4x}\right)-
\frac{(1-x)^2}{4x}\log_2 \left(\frac{(1-x)^2}{4x}\right) \, .
\]
Such a quantity is, again, a
decreasing function of $\tilde{\nu}_{-}$, thus providing
a quantification of the entanglement of symmetric states equivalent
to the one provided by the logarithmic negativity $E_{\N}$.

In the nonsymmetric case, an important result is that for any
entangled two--mode Gaussian state $\varrho$, the symplectic
eigenvalue $\tilde{\nu}_{-}$ (and, consequently, the logarithmic
negativity) can be estimated with remarkable accuracy by only
determining the global purity $\tr{\varrho^2}$ of the state, and
the two local purities $\tr{\varrho_{1,2}^2}$ of each of the two
reduced single--mode states $\varrho_i = {\rm Tr}_j \varrho$. For
the aforementioned class of nonsymmetric thermal squeezed states,
the estimate becomes actually an exact quantification, since the
logarithmic negativity of these states is a function of the three
purities only \cite{asiprl}. These states are indeed the maximally
entangled two--mode Gaussian states at fixed global and local
purities, and thus they are the states one would like to produce
and exploit in any continuous-variable quantum information
processing. On the other hand, the symmetric instance, which
carries the highest possible entanglement \cite{extremal} (and
which is the experimental product of the present paper), is the
one that enables continuous-variable teleportation of an unknown
coherent state \cite{teleportationkimble,australie,furusawa} with
a fidelity arbitrarily close to 1 even in the presence of noise
(mixedness), provided that the state is squeezed enough (ideally,
a unit fidelity requires infinite squeezing). Actually, the
fidelity of such an experiment, if the squeezed thermal states
employed as shared resource are optimally produced, turns out to
be itself a measure of entanglement and provides a direct,
operative quantification of the entanglement of formation present
in the resource \cite{telepoppate}.

\section{Effects of mode coupling on the entanglement generation}
\label{sec:modecoupling}

As mentioned in the introduction, entanglement is very often
produced by mixing on a beam splitter two squeezed modes. In the
general case, the squeezed quadratures have an arbitrary phase
difference. We denote $\theta+\pi/2$ the phase difference between
the two squeezed quadratures. The CM of the squeezed modes is then
\begin{equation}
 \Gamma_{A_{+}\,\!A_{-}} = \left(
\begin{array}{cc|cc}
a & 0 & 0 & 0 \\
0 & 1/a & 0 & 0 \\
\hline 0 & 0 & b & c\\
0 & 0 & c & b'
\end{array} \right), \label{eq:sqnondiag}
\end{equation}
while the CM of the two modes after the beam-splitter is
\begin{equation}
\Gamma_{A_{1}\,\!A_{2}} = R_{\pi/4}^T .  \Gamma_{A_{+}\,\!A_{-}} .
R_{\pi/4} = \left(
\begin{array}{cc|cc}
n_1 & k' & k & k' \\
k' & n_2 & k' & -k \\
\hline k & k' & n_1 & k'\\
k' & -k & k' & n_2
\end{array} \right)\label{eq:nonsform}
\end{equation}
where
\begin{eqnarray*}
  b\!\!&=&\!\! \frac{\cos^2 \theta}a + a \sin^2 \theta\,,\quad
  b'= a \cos^2 \theta + \frac{\sin^2 \theta}a\,,\quad
  c = \left(a - \frac1a\right) \sin\theta \cos\theta\,,\\
  n_1 \!\!&=&\!\! \frac{\cos^2\theta + a^2 (\sin^2\theta+1)}{2a}\,,\quad
  n_2 = \frac{a^2\cos^2\theta +
\sin^2\theta+1}{2a}\,, \\
  k\!\!&=&\!\! \left(\frac{1-a^2}{2a}\right) \cos^2\theta\,,\quad
  k' = \left(\frac{a^2-1}{2a}\right) \sin\theta \cos\theta\,.\\
 \end{eqnarray*}
Let us first note that expression \ref{eq:nonsform} can be brought
back to the expression given on fig. 5 of \cite{eprparis}
\emph{via} local unitary operations which do not modify the
entanglement.

The CM of the squeezed ($A_\pm$) modes gives a good insight into
the properties of the two-mode state. One can see that the
intermodal blocks are zero, meaning that the two modes are
uncorrelated. Consequently, they are the two most squeezed modes
of the system (no further passive operation can select more
squeezed quadratures). But one can also note that the two diagonal
blocks are not diagonal simultaneously. This corresponds to the
tilt angle of the squeezed quadrature. In order to maximize the
entanglement, the two squeezed quadratures have to be made
orthogonal, which can be done by a phase-shift of one mode
relative to the other.

\begin{figure}[h!]
\centering{\includegraphics[width=.75\columnwidth]{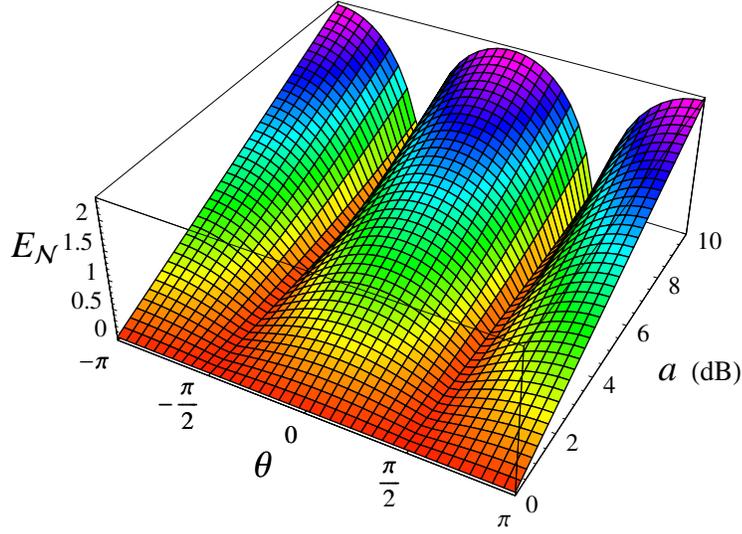}%
\caption{\label{loneper}Logarithmic negativity as a function of
the single--mode squeezing $a$ and the tilt angle $\theta$ between
the two non-orthogonal quadratures in presence of mode coupling.}}
\end{figure}

It is easy in fact to compute the logarithmic negativity
quantifying entanglement between the entangled modes $A_1$ and
$A_2$, when the two squeezed quadratures are rotated of $\pi/2 +
\theta$. One has $E_\N (\Gamma_{A_{1}\,\!A_{2}}) = -(1/2) \log
\tilde \nu ^2$, with
\begin{eqnarray}
\tilde\nu^2 &=& \left(\frac{1}{4a^2}\right)\Bigg\{2 \left(a^4 +
1\right) \cos ^2(\theta ) + 4 a^2 \sin ^2(\theta ) \nonumber \\
&-& \sqrt{2} \left(a^2 -1\right) \sqrt{\cos ^2(\theta) \left[a^4 +
6 a^2 + \left(a^2 - 1\right)^2 \cos (2 \theta ) +
1\right]}\Bigg\}\,. \label{eq:nu}
\end{eqnarray}
The symplectic eigenvalue $\tilde\nu$ is obviously a periodic
function of $\theta$, and it is globally minimized for $\theta =
k\,\pi$, with $k\in {\mathbbm Z}$. The entanglement, in other
words, is maximized for orthogonal modes in phase space, as
already predicted in Ref. \cite{passive}. Notice that this results
holds for general nonsymmetric states, i.e. also in the case when
the two modes $A_1$ and $A_2$ possess different individual
squeezings. For symmetric states, the logarithmic negativity is
depicted as a function of the single--mode squeezing $a$ and the
tilt angle $\theta$ in figure \ref{loneper}.

\begin{figure}[h!]
\centering{\includegraphics[width=.95\columnwidth]{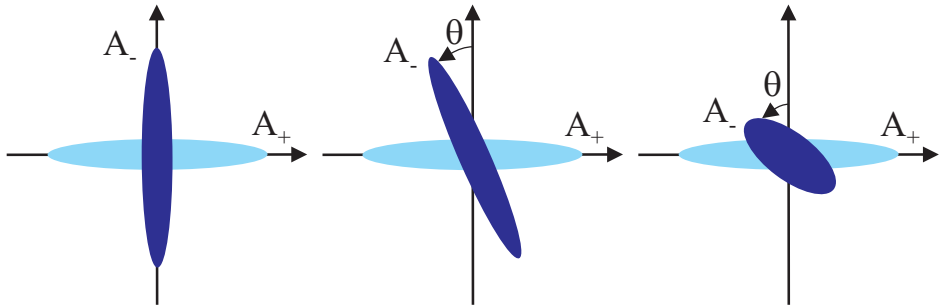}%
\caption{\label{ellipse}Fresnel representation of the noise
ellipse of the $\pm 45^{\circ}$ rotated modes when the coupling is
increased. The noise ellipse of the $-45^{\circ}$ mode rotates and
the noise reduction is degraded when the coupling increases while
the $+45^{\circ}$ rotated mode is not affected.}}
\end{figure}

In the experiment we will discuss below, the entanglement is
produced by a single device, a type-II OPO operated below
threshold. When no coupling is present in the optical cavity, the
entangled modes are along the neutral axis of the crystal while
the squeezed modes corresponds to the $\pm 45^\circ$ linear
polarization basis. However, we have shown theoretically and
experimentally in \cite{eprparis} that a coupling can be added
\emph{via} a birefringent plate which modifies the quantum
properties of this device: the most squeezed quadratures are
non-orthogonal with an angle depending on the plate angle. When
the plate angle increases, the correlated ($A_-$) quadrature
rotates of a tilt angle $\theta$ and the correlations are
degraded. The evolution is depicted on figure \ref{ellipse}
through the noise ellipse of the superposition modes.

Eqn. \ref{eq:nu} shows that when coupling is present, it is
necessary to perform an operation on the two modes in order to
optimize the available entanglement. Before developing
experimental measures of entanglement and optimization of the
available resource in our system, let us detail our experimental
setup.

\section{Experimental setup and homodyne measurement}
\label{sec:setup}

\begin{figure*}
\centering{\includegraphics[width=.95\columnwidth]{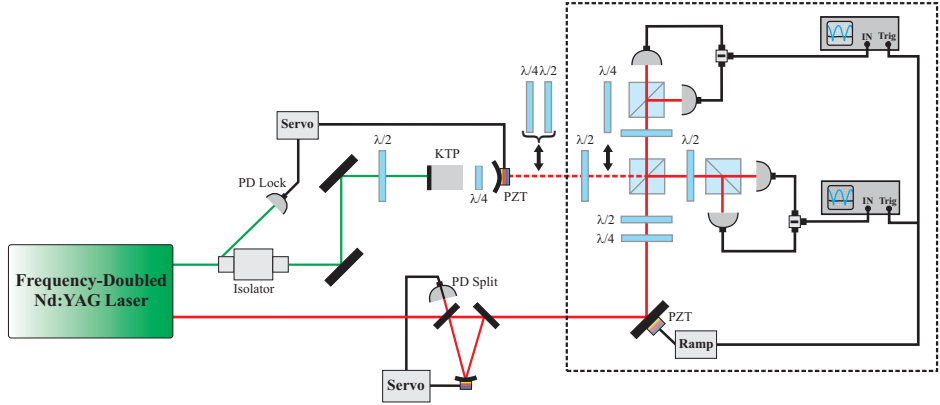}
\caption{Experimental setup. A continuous-wave frequency-doubled
Nd:YAG laser pumps below threshold a type II OPO with a $\lambda/4$
plate inserted inside the cavity. The generated two-mode vacuum
state is characterized by two simultaneous homodyne detections. The
infrared output of the laser is used as local oscillator after
filtering by a high-finesse cavity. The two couples
$\{\lambda/4,\lambda/2\}$ on each path are used to make arbitrary
phase shift between orthogonal components of polarization. PD Lock:
FND-100 photodiode for locking of the OPO. PD Split: split
two-element InGaAs photodiode for tilt-locking of the filtering
cavity.}\label{setup}}
\end{figure*}

The experimental scheme is depicted on figure \ref{setup} and
relies on a frequency-degenerate type-II OPO below threshold. The
system is equivalent to the one of the seminal experiment by Ou
\emph{et al.} but a $\lambda/4$ birefringent plate has been
inserted inside the optical cavity. When this plate is rotated, it
results in a linear coupling between the signal and idler modes
which induces above threshold a phase locking effect at exact
frequency degeneracy \cite{wong98,opophlockc}. This
triply-resonant OPO is pumped below threshold with a continuous
frequency-doubled Nd$:$YAG laser. The input flat mirror is
directly coated on one face of the 10mm-long KTP crystal. The
reflectivities for the input coupler are 95\% for the pump (532nm)
and almost 100\% for the signal and idler beams (1064nm). The
output coupler (R=38mm) is highly reflecting for the pump and its
transmission is 5\% for the infrared. At exact triple resonance,
the oscillation threshold is less than 20 mW. The OPO is actively
locked on the pump resonance by the Pound-Drever-Hall technique.
The triple resonance is reached by adjustment of both the crystal
temperature and the frequency of the pump laser. Under these
conditions, the OPO can operate stably during more than one hour
without mode-hopping. The birefringent plate inserted inside the
cavity is exactly $\lambda/4$ at 1064 nm and almost $\lambda$ at
the 532 nm pump wavelength. Very small rotations of this plate
around the cavity axis can be operated thanks to a piezoelectric
actuator.

Measurements of the quantum properties of arbitrary quadratures of
light mode are generally made using homodyne detection
\cite{yuen83}. When an intense local oscillator is used, one
obtains a photocurrent which is proportional to the quantum noise
of the light in a quadrature defined by the phase-shift between
the local oscillator and the beam measured. This photocurrent can
be either sent to a spectrum analyzer which calculates the noise
power spectrum or numerized for further treatments like
tomographic measurements of the Wigner function \cite{tomography}
or selection \cite{laurat03}. As mentioned above, one can also
characterize the entanglement by looking at linear combinations of
the optical modes as opposed to linear combinations of the
photocurrents \cite{duan00,simon00}. The two-modes which form the
entangled state must be transformed via the beam splitter
transformation, that is they are mixed on a 50/50 beam splitter or
a polarizing beam-splitter into two modes which will be both
squeezed if the original state is entangled.

Homodyne detection allows for a simple and direct measurement of
the $2\times 2$ diagonal blocks of the $4\times 4$ CM. In order to
measure the $2\times 2$ off-diagonal blocks, linear combinations
of the photocurrents can be used. One can thus obtain a complete
measurement of the two-mode CM. However, in order to characterize
simultaneously two modes a single phase reference is needed. To
implement this, we have built a simultaneous double homodyne
detection (Fig. \ref{setup}, in box). The difference photocurrents
are sent into two spectrum analyzers triggered by the same signal.
Two birefringent plates inserted in the local oscillator path are
rotated in order to compensate residual birefringence. A
$\lambda/4$ plate can be added on the beam exiting the OPO in
order to transform the in-phase detections into in-quadrature ones
(making the transformation $(\hat x_1, \hat p_1, \hat x_2 \hat
p_2) \ensuremath{\to} (\hat x_1, \hat p_1, \hat p_2 \hat x_2)$).
In such a configuration, two states of light with squeezing on
orthogonal quadratures give in phase squeezing curves on the
spectrum analyzers.

\section{Experimental measures of entanglement by the negativity}
\label{sec:expmesent}

\begin{figure}[t!]
\includegraphics[width=13cm]{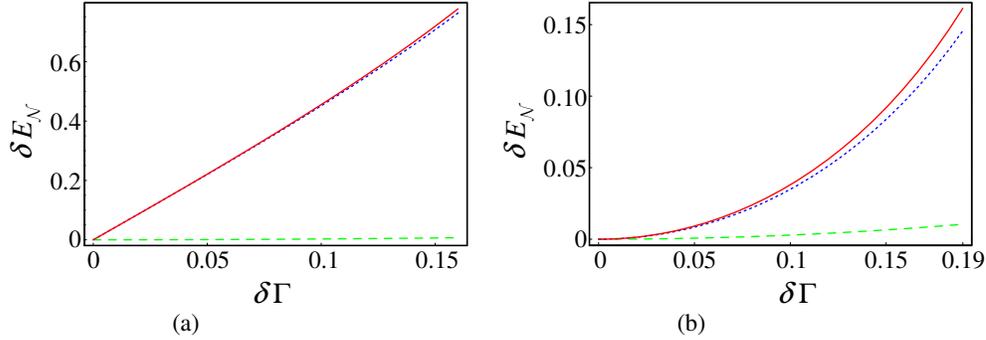}
\caption{Error $\delta E_{\mathcal N}$ on the logarithmic negativity
between modes $A_1$ and $A_2$, as a function of the error
$\delta\Gamma$ on the entries of the diagonal (a) and off-diagonal
(b) $2\times 2$ blocks of the measured CM $\Gamma$ in the $A_\pm$
basis, given by \eq{exem}. In plot (a): the solid red curve refers
to equal errors (of value $\delta\Gamma$) on the eight entries of
the diagonal blocks, the dotted blue curve refers to equal errors on
the four diagonal entries of the diagonal blocks while the dashed
green curve refers to equal errors on the off-diagonal entries of
the diagonal blocks (non standard form entries). At
$\delta\Gamma\gtrsim0.16$ some of the considered states get
unphysical. In plot (b): the solid red curve refers to equal errors
on the four entries of the off-diagonal block, the dotted blue curve
refers to equal errors on the two off-diagonal entries of the
off-diagonal block (non standard form entries), while the dashed
green curve refers to equal errors on the diagonal entries of the
off-diagonal block (standard form entries). At
$\delta\Gamma\gtrsim0.19$ some of the considered states get
unphysical. \label{error}}
\end{figure}

As all experimental measurements, the measurement of the CM  is
subject to noise. It is thus critical to evaluate the influence of
this noise on the entanglement. A quantitative analysis, relating
the errors on the measured CM entries (in the $A_\pm$ basis) to
the resulting error in the determination of the logarithmic
negativity (the latter quantifying entanglement between the
corresponding $A_1$ and $A_2$ modes) has been carried out and is
summarized in Fig.~\ref{error} in absence of mode coupling. In
general, the determination of the logarithmic negativity is much
more sensitive to the errors on the diagonal $2\times 2$ blocks
$\alpha$ and $\beta$ (referring to the reduced states of each
mode, see \eq{espre}) of the CM $\Gamma$ than on the off-diagonal
ones ($\gamma$, and its transpose $\gamma^T$, whose expectations
are assumed to be null). Let us remark that the relative stability
of the logarithmic negativity with respect to the uncertainties on
the off-diagonal block adds to the reliability of our experimental
estimates of the entanglement. Notice also that, concerning the
diagonal blocks, the errors on diagonal (standard form) entries
turn out to affect the precision of the logarithmic negativity
more than the errors on off-diagonal (non standard form) entries.
This behavior is reversed in the off-diagonal block, for which
errors on the off-diagonal (non standard form) entries affect the
uncertainty on the entanglement more than errors on the diagonal
(standard form) entries.

Experimentally, we have measured the noise on the CM elements to
be at best on the order of a few percents of the measured values
for the diagonal blocks, corresponding to a fraction of a dB
\cite{db}. This is the case for the diagonal blocks which are
well-known since they are directly related to the noise
measurements of $A_+$ and $A_-$. The situation is less favorable
for the off-diagonal blocks: the off-diagonal elements of these
blocks show a large experimental noise leading which, as shown on
Fig. \ref{error} (b), may lead in some cases to unphysical CM,
yielding for instance a negative determinant and complex values
for the logarithmic negativity. In the following, we will set
these terms to zero in agreement with the form of the CM of
Eq.~\ref{eq:sqnondiag}.

\begin{figure}[htpb!]
\centering{\includegraphics[width=.85\columnwidth,clip=]{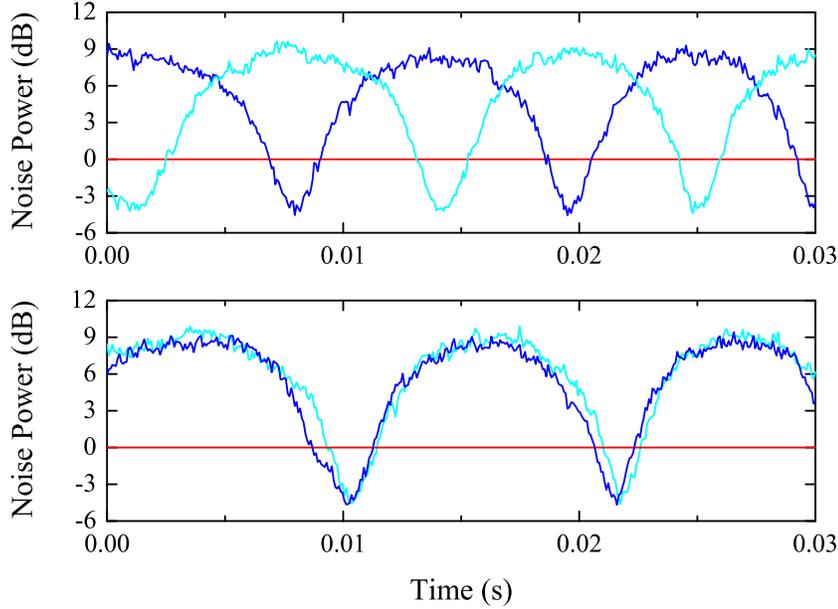}%
\caption{\label{scan}Normalized noise variances at 3.5 MHz of the
$\pm 45^{\circ}$ modes while scanning the local oscillator phase.
The first plot corresponds to in-phase homodyne detections and the
second one in-quadrature. Squeezing is well observed on orthogonal
quadratures. (RBW 100 kHz, VBW 1 kHz)}}
\end{figure}

Let us first give an example of entanglement determination from
measurements of CM elements, in the absence of mode coupling.
Without the plate, the squeezing of the two superposition modes is
expected on orthogonal quadratures (the ideal CM is then in the
form \ref{eq:sqnondiag} with $\theta=0$). Spectrum analyzer traces
while scanning the local oscillator phase are shown on figure
\ref{scan}: the rotated modes are squeezed on orthogonal
quadratures. The state is produced directly in the standard form
and the CM in the $A_\pm$ basis can be extracted from this
measurement:
\begin{equation} \label{exem}
\Gamma (\rho=0) = \left(\begin{array}{cc|cc}
0.33 & 0 & (0) & (0)\\
0 & 7.94 & (0) & (0) \\ \hline
(0) & (0) & 7.94 & 0\\
(0) & (0) & 0 & 0.33
\end{array}\right)
\end{equation}
The resulting smallest symplectic eigenvalue is the geometric
average of the two minimal diagonal elements : $\tilde{\nu}_- =
0.33$, yielding a logarithmic negativity $E_{\N} = - \log_2
(\tilde{\nu}_-) = 1.60$ between the modes $A_1$ and $A_2$.

\section{Experimental non standard form and optimization by linear optics}
\label{sec:nonsf}

\begin{figure}[t!]
\centering{\includegraphics[width=.85\columnwidth,clip=]{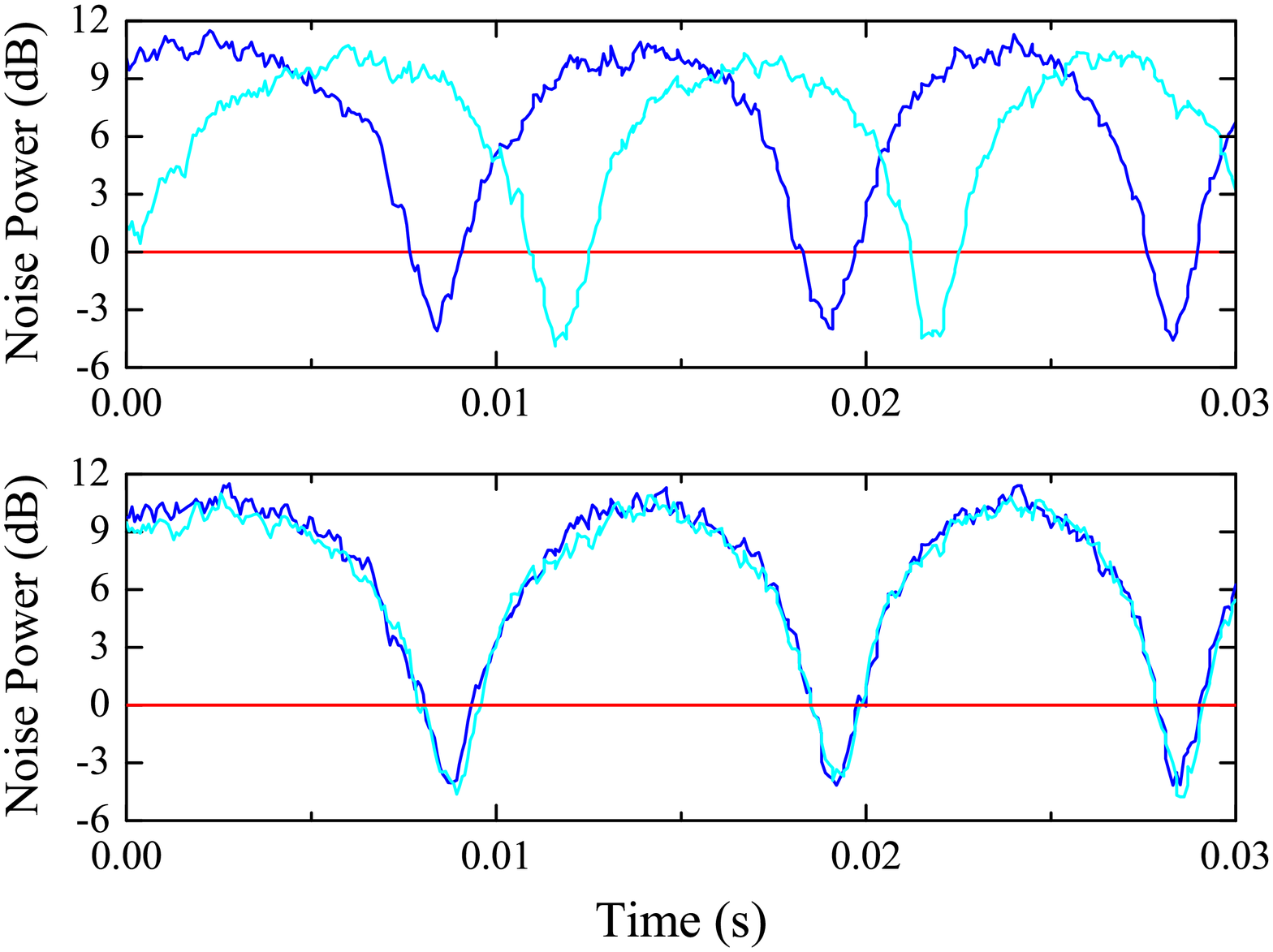}%
\caption{\label{corrige} Normalized noise variances at 3.5 MHz of
the rotated modes while scanning the local oscillator phase for an
angle of the plate of $0.3^{\circ}$, before and after the
non-local operation. The homodyne detections are in-quadrature.
After this operation, squeezing is observed on orthogonal
quadratures.}}
\end{figure}

As discussed previously, when the plate angle is increased, the
state produced is not anymore in the standard form but rather
similar to eqn. \ref{eq:sqnondiag}. Figure \ref{corrige} gives the
normalized noise variances at 3.5~MHz of the rotated modes while
scanning the local oscillator phase for an angle of the plate of
$0.3^{\circ}$. The first plot shows that the squeezing is not
obtained on orthogonal quadratures. The CM takes the following
form:
\begin{equation}
\Gamma (\rho=0.3^{\circ}) = \left(\begin{array}{cc|cc}
 0.4 & 0 & (0) & (0)\\
0 & 12.59 & (0) & (0) \\ \hline
(0) & (0) & 9.54 & -5.28\\
(0) & (0) & -5.28 & 3.45
\end{array}\right)
\end{equation}
In this instance, the logarithmic negativity between $A_1$ and $A_2$
is much lower than the previous value: $E_{\N}  = 1.13$. We will now
study the action of passive operations on this two-mode state.

As stated above, entanglement can be increased \emph{via} passive
operations performed simultaneously on the two modes. Such
operations include phase-shifts and beam splitters, which can be
readily performed on co-propagating, orthogonally polarized beams
\cite{simon}. The minimal combination of waveplates can be shown
to consist in three waveplates : two $\lambda/4$ waveplates
denoted Q and one $\lambda/2$ waveplate denoted H. When using any
combination of these three plates, the state can be put back into
standard form which will maximize the entanglement. This operation
consists in a phase-shift of the rotated modes. Figure
\ref{corrige} gives the normalized noise variances before and
after this operation. The CM is changed to:
\begin{equation}
\Gamma (\rho=0.3^{\circ}) = \left(\begin{array}{cc|cc}
 0.4 & 0 & (0) & (0)\\
0 & 12.59 & (0) & (0) \\ \hline
(0) & (0) & 12.59 & 0\\
(0) & (0) & 0 & 0.4
\end{array}\right) \, ,
\end{equation}
giving a logarithmic negativity $E_{\N} = 1.32$ between $A_1$ and
$A_2$, larger than the value before the operation. It is also the
maximal value than can be obtained considering the available
entanglement.

Let us remark again that this transformation is non-local in the
sense of the EPR argument: it has to be performed before spatially
separating the entangled modes for a quantum communication
protocol for instance.

\section{Conclusion}

We have described the powerful tools underlying the description of
continuous variables systems in quantum optics. These tools allow
for a nice pictorial view of two-mode Gaussian entangled states.
Specifically, we have illustrated their properties through the
description and manipulation of the entanglement produced
experimentally with an original device, a type-II optical
parametric oscillator containing a birefringent plate. We have
demonstrated the capabilities of this system through entanglement
measurements and manipulation of entangled states, showing, in
particular, how entanglement can be maximized using purely passive
operations.

We have also studied quantitatively the influence of the noise,
affecting the measurement of the elements of the CM, on the
entanglement, showing that the most significant covariances
(exhibiting the highest stability against noise) for an accurate
entanglement quantification are the diagonal terms of the diagonal
single-mode blocks, and the off-diagonal terms of the intermodal
off-diagonal block, the latter being the most difficult to measure
with high precision. Alternative methods have been devised to
tackle this problem \cite{fiurasek,asiprl} based on direct
measurements of global and local invariants of the CM. Such
techniques have been implemented in the case of pulsed beams
\cite{dethompuls} but no experiment to date has been performed for
continuous-wave beams.

\section*{Acknowledgement}
Laboratoire Kastler-Brossel, of the Ecole Normale Sup\'{e}rieure
and the Universit\'{e} Pierre et Marie Curie, is associated with
the Centre National de la Recherche Scientifique (UMR 8552).
Laboratoire Mat{\'e}riaux et Ph{\'e}nom{\`e}nes Quantiques, of Universit\'e
Denis Diderot, is associated with the Centre National de la
Recherche Scientifique (UMR 7162). This work has been supported by
the European Commission project QUICOV (IST-1999-13071) and ACI
Photonique (Minist\`ere de la Recherche et de la Technologie).
JAOH acknowledges financial support from CAPES/COFECUB. AS
acknowledges financial support from EPSRC through the grant
QIP-IRC. GA and FI acknowledge INFM, INFN, and MIUR under national
project PRIN-COFIN 2002 for financial support.


\begin{thebibliography}{99}
\bibitem{abs}{Here $|A|$ stands for the
absolute value of the diagonalizable matrix $A$, to be understood in the usual operatorial sense:
if $A=T^{-1}DT$, where $D$ is diagonal and $T$ is a diagonalizing operator, then
$|A|=T^{-1}|D|T$, where $|D|$ is just the diagonal matrix with the absolute values of the
eigenvalues of $A$ as entries.}
\bibitem{db} The noise expressed in decibels (dB) is obtained from
the covariance matrix elements via the formula $N_{ij}(dB) = 10
\log_{10}(\Gamma_{ij})$
\end{thebibliography}
\end{document}